\documentclass[conference,twocolumn]{IEEEtran}

\usepackage{graphicx}
\usepackage{color}
\usepackage{placeins}
\usepackage{float}
\usepackage{tabularx,colortbl}
\usepackage{amssymb}
\usepackage{amsthm}
\usepackage{cite}
\usepackage{amsmath}

\usepackage{caption2}
\theoremstyle{plain}
\newtheorem{thm}{Theorem}

\theoremstyle{plain}
\newtheorem{rem}{Remark}

\IEEEoverridecommandlockouts

\begin{document}

\title{Performance Analysis of Cell-Free Massive MIMO Systems with Asynchronous Reception}

\author{Jiakang Zheng, Zhuoyi Zhao, Jiayi Zhang, Julian Cheng, and Victor C. M. Leung

\thanks{J. Zheng, Z. Zhao, and J. Zhang, are with the School of Electronic and Information Engineering, Beijing Jiaotong University, Beijing 100044, China (e-mail: \{jiakangzheng, jiayizhang\}@bjtu.edu.cn).}
\thanks{J. Cheng is with the School of Engineering, The University of British Columbia, Kelowna, BC V1V 1V7, Canada (e-mail: julian.cheng@ubc.ca).}
\thanks{V. C. M. Leung is with the College of Computer Science and Software Engineering, Shenzhen University, Shenzhen 518060, China, and also with the Department of Electrical and Computer Engineering, The University of British Columbia, Vancouver, BC V6T 1Z4, Canada (e-mail: vleung@ece.ubc.ca).}
}

\maketitle
\vspace{-1cm}
\begin{abstract}
Cell-free (CF) massive multiple-input multiple-output (MIMO) is considered as a promising technology for achieving the ultimate performance limit. However, due to its distributed architecture and low-cost access points (APs), the signals received at user equipments (UEs) are most likely asynchronous.
In this paper, we investigate the performance of CF massive MIMO systems with asynchronous reception, including both effects of delay and oscillator phases. Taking into account the imperfect channel state information caused by phase asynchronization and pilot contamination, we obtain novel and closed-form downlink spectral efficiency (SE) expressions with coherent and non-coherent data transmission schemes, respectively. Simulation results show that asynchronous reception destroys the orthogonality of pilots and coherent transmission of data, and thus results in poor system performance. In addition, getting a highly accurate delay phase is substantial for CF massive MIMO systems to achieve coherent transmission gain. Moreover, the oscillator phase of UEs has a larger effect on SE than that of the APs, because the latter can be significantly reduced by increasing the number of antennas.
\end{abstract}


\IEEEpeerreviewmaketitle

\section{Introduction}

Cell-free (CF) massive multiple-input multiple-output (MIMO) has been envisioned to provide uniform spectral efficiency (SE) and ubiquitous connectivity for six-generation (6G) wireless network \cite{zhang2020prospective,9205230,guo2021beam}. The key idea of CF massive MIMO systems is that through spatial multiplexing on the same time-frequency resource, many geographically distributed access points (APs) connected to a central processing unit (CPU) serve the user equipments (UEs) coherently \cite{Ngo2017Cell}. Therefore, the huge macro-diversity gain of CF massive MIMO systems brought by the joint transmission and reception has made academic interest in this area grow exponentially \cite{chen2021survey}. For instance, in terms of 95\%-likely per-user SE, CF massive MIMO systems outperform small-cell systems due to the joint interference cancellation capability \cite{bjornson2019making}. In addition, compared to traditional cellular systems, the CF massive MIMO systems based joint signal processing can effectively alleviate the influence of non-ideal factors, such as channel aging, hardware impairments, etc. \cite{9322468,zheng2021uav,zhao2021application}.

However, most current works on CF massive MIMO assume perfect synchronization to ensure the joint coherent processing is feasible, but this is impractical in a distributed architecture \cite{zhang2021improving}. One reason is that the geographically distributed APs cause unavoidable differences in the arrival time of the signal at the UE \cite{8676341}. The resulting delay phases are multiplied in the received signals, raising a challenge for CF massive MIMO systems to achieve coherent transmission \cite{6760595}. Another reason is that the transmitter and receiver hardware are not reciprocal in that they can cause different phase shifts on the channels \cite{6902790}. Besides, this multiplicative oscillator phase changes gradually at each channel use, and is modeled by the Wiener process \cite{bjornson2015massive}.
Generally speaking, the asynchronous reception in CF massive MIMO mainly arises from two factors: the delay phase caused by the distributed architecture and the oscillator phase caused by low-cost hardware. These impurities lead to a loss in the achievable signal-to-noise ratio (SNR), and seriously affect the practical deployment of CF massive MIMO systems \cite{9531352,9502552}.
Therefore, technical analysis to assess the impacts of asynchronous reception on CF massive MIMO arises as an interesting research subject.

Motivated by the foregoing observations, we analyze the performance of CF massive MIMO systems with asynchronous reception, including both delay and oscillator phases. Considering the imperfect channel state information from pilot contamination and phase asynchronization, we obtain closed-form downlink SE expressions for coherent and non-coherent data transmission schemes, respectively. It is reveal that asynchronous reception destroys the orthogonality of pilots and the coherent transmission of data, and thus leads to a poor SE performance of CF massive MIMO systems.



\section{System Model}\label{se:model}

In this paper, we study a CF massive MIMO system comprising $L$ APs and $K$ UEs. Besides, one single antenna and $N$ antennas are deployed for each UE and AP, respectively. It is assumed that all $L$ APs simultaneously serve all $K$ UEs on the same time-frequency resource \cite{Ngo2017Cell}. Moreover, we use the time-division duplex protocol with a standard coherence block model consisting of $\tau_c$ time instants (channel uses), with the uplink training phase occupying $\tau_p$ time instants and the downlink transmission phase occupying $\tau_c - \tau_p$ time instants. Besides, the frequency-flat channel between AP $l$ and UE $k$ at each coherence block is modeled as Rayleigh fading \cite{bjornson2019making}:
\begin{align}
{{\mathbf{h}}_{kl}} \sim \mathcal{C}\mathcal{N}\left( {{\mathbf{0}},{{\mathbf{R}}_{kl}}} \right)
\end{align}
where ${{\mathbf{R}}_{kl}} \in {\mathbb{C}^{N \times N}}$ represents the spatial correlation matrix and ${\beta _{kl}} \!\triangleq\! {\text{tr}}\left( {{{\mathbf{R}}_{kl}}} \right)/N$ means the large-scale fading coefficient.

\subsection{Asynchronous Reception}


The asynchronous reception of the transceiver comes mainly from two factors: propagation delay difference and hardware oscillator error. Specifically, the channel is multiplied by the caused delay and oscillator phases.

\subsubsection{Delay Phase}

Because of the different positions of the APs in the CF architecture, the distances between each AP and a certain UE are different, resulting in asynchronous signal arrival to UEs. This asynchronous reception effect will introduce a phase shift as \cite{8676341}
\begin{align}
{\theta _{kl}} = {e^{ - j2\pi \frac{{{\Delta t_{kl}}}}{T_s}}}
\end{align}
where $\Delta t_{kl} = \Delta d_{kl}/c$ is the timing offset of the signal intending for the $k$th UE and transmitted by the $l$th AP. Besides, $\Delta d_{kl}$, $c$, and $T_s$ are the propagation delay, speed of light, and symbol duration, respectively. Without loss of generality, we assume that the first arrived signal to UE $k$ is from AP ${l'}$ and its timing offset is $\Delta t_{kl'} = 0$.
\subsubsection{Oscillator Phase}

Each AP and UE are assumed to have their own free-running oscillator, and the phase of transmission symbol in each channel uses changes due to the phase noise. Then, the oscillator phase between AP $l$ and UE $k$ at each time instant can be defined by ${\vartheta _{kl}}\left[ n \right] \triangleq \exp\left({\varphi _k}\left[ n \right] + {\phi _l}\left[ n \right]\right)$ with the discrete-time Wiener phase model \cite{bjornson2015massive}
\begin{align}
{\varphi _k}\left[ n \right] &= {\varphi _k}\left[ {n - 1} \right] + \delta _k^{{\text{ue}}}\left[ n \right] \\
{\phi _l}\left[ n \right] &= {\phi _l}\left[ {n - 1} \right] + \delta _l^{{\text{ap}}}\left[ n \right]
\end{align}
where ${\varphi _k}\left[ n \right]$ and ${\phi _l}\left[ n \right]$ are the oscillator phase of UE and AP at the $n$th time instant, respectively. Besides, $\delta _k^{{\text{ue}}}\left[ n \right] \sim \mathcal{C}\mathcal{N}\left( {0,\sigma _{{k}}^2} \right)$ and $\delta _l^{{\text{ap}}}\left[ n \right] \sim \mathcal{C}\mathcal{N}\left( {0,\sigma _{{l}}^2} \right)$ are the phase increment of AP and UE at the $n$th time instant. Note that $\sigma _i^2 = 4{\pi ^2}{f_c^2}{c_i}{T_s},i = {k}, {l} $ denote the phase increment variance, where $f_c$ is the carrier frequency and $c_i$ is a constant dependent on the oscillator.
\begin{rem}
In our analysis, we focus on the scenario, where each AP and UE has its oscillator so that their oscillator phase processes are considered mutually independent. But in all cases, we assume independent and identically distributed oscillator phase statistics across different APs and UEs, i.e., $\sigma _{{k}}^2 = \sigma _{\mathrm{ue}}^2$ and $\sigma _{{l}}^2 = \sigma _{\mathrm{ap}}^2$, $\forall k,l$.
\end{rem}
Considering the effect of both delay and oscillator phases, the channel between the $k$th UE and the $l$th AP at the $n$th time instant is expressed as
\begin{align}\label{gkl}
{{\mathbf{g}}_{kl}}\left[ n \right] = {\theta _{kl}}{{\mathbf{h}}_{kl}}\left[ n \right] = {\theta _{kl}}{\vartheta _{kl}}\left[ n \right]{{\mathbf{h}}_{kl}} ,n = 1, \ldots ,{\tau _c}
\end{align}
where ${{\mathbf{h}}_{kl}}\left[ n \right]$ is the channel that combines the oscillator phase and it is random in each time instant. Besides, ${\theta _{kl}}$ is mainly determined by the positions of APs and UEs, and thus can be considered as a constant among multiple coherent blocks\footnote{We assume that the delay phase can be perfectly known by positioning or other technologies. However, we will use the downlink precoding with or without (used/forgotten) delay phase to quantifying its impact \cite{9531352}.}.

\subsection{Channel Estimation}

We employ $\tau_p$ mutually orthogonal time-multiplexed pilot sequences, which implies that pilot sequence $t$ equates to transmitting a pilot signal only at the $t$th time instant. Besides, a huge network with $K>\tau_p$ is studied, in which different UEs use the same time instant. Moreover, the index of the time instant allocated to UE $k$ is represented by ${t_k} \in \left\{ {1, \ldots ,{\tau _p}} \right\}$, the other UEs that use the same time instant for pilot transmission as UE $k$ is defined by ${\mathcal{P}_k} = \left\{ {i:{t_i} = {t_k}} \right\} \subset \{ 1, \ldots ,K\} $. Considering the effect of asynchronous reception, the received signal between AP $l$ and UE $k$ at time instant $t_k$ is given by
\begin{align}\label{zl1}
{{\mathbf{z}}_l}\left[ {{t_k}} \right] = \sum\limits_{i \in {\mathcal{P}_k}} {\sqrt {{p_i}} {{\mathbf{g}}_{il}}\left[ {{t_i}} \right] + {{\mathbf{w}}_l}\left[ {{t_k}} \right]}
\end{align}
where $p_i \geqslant 0$ denotes the power of pilot transmitted from UE $i$ and ${{\mathbf{w}}_l}\left[ {{t_k}} \right] \sim \mathcal{C}\mathcal{N}\left( {0,{\sigma ^2}{{\mathbf{I}}_N}} \right)$ represents the receiver noise.
This received signal can be used to estimate (or predict) channel realization at any time instant in the block, but the accuracy of the estimate degrades with the temporal gap between the considered channel realization and the pilot transmission grows. We then consider the estimates at the channels at time instant $\tau_p + 1$ without losing generality. In addition, $\lambda = \tau_p + 1$ is defined to simplify the notation, we can write \eqref{zl1} as
\begin{align}
{{\mathbf{z}}_l}\left[ {{t_k}} \right] &= \sqrt {{p_k}} {\theta _{kl}}\Theta _{kl}^*\left[ \lambda - t_k  \right]{{\mathbf{h}}_{kl}}\left[ \lambda  \right] \notag\\
&+ \sum\limits_{i \in {\mathcal{P}_k}/\left\{ k \right\}} {\sqrt {{p_i}} {\theta _{il}}{{\mathbf{h}}_{il}}\left[ {{t_i}} \right] + {{\mathbf{w}}_l}\left[ {{t_k}} \right]}
\end{align}
where
\begin{align}\label{theta1}
{\Theta _{kl}}\left[ \lambda - t_k \right] &= {\vartheta _{kl}}\left[ \lambda \right]\vartheta _{kl}^*\left[ t_k  \right] \notag\\
 &= \exp \left( {j\sum\limits_{s = t_k  + 1}^\lambda {\left( {\delta _k^{{\text{ue}}}\left[ s \right] + \delta _l^{{\text{ap}}}\left[ s \right]} \right)} } \right) .
\end{align}
By the characteristic function of Gaussian random variable, the mean of \eqref{theta1} is given by
\begin{align}\label{theta2}
\mathbb{E}\left\{ {{\Theta _{kl}}\left[ \lambda - t_k \right]} \right\} = {e^{ - \frac{{\lambda  - {t_k}}}{2}\left( {\sigma _{{\text{ap}}}^2 + \sigma _{{\text{ue}}}^2} \right)}} .
\end{align}
Then, using standard minimum mean square error (MMSE) estimation \cite{Ngo2017Cell}, the MMSE estimate ${{{\mathbf{\hat h}}}_{kl}}\left[ \lambda  \right]$ of the channel coefficient ${{\mathbf{h}}_{kl}}\left[ \lambda  \right]$ can be computed by each AP $l$ as
\begin{align}\label{hhat}
{{{\mathbf{\hat h}}}_{kl}}\left[ \lambda  \right] = \sqrt {{p_k}} {e^{ - \frac{{\lambda  - {t_k}}}{2}\left( {\sigma _{{\text{ap}}}^2 + \sigma _{{\text{ue}}}^2} \right)}}\theta _{kl}^{\text{*}}{{\mathbf{R}}_{kl}}{{\mathbf{\Psi }}_{kl}}{{\mathbf{z}}_l}\left[ {{t_k}} \right]
\end{align}
where
\begin{align}
{{\mathbf{\Psi }}_{kl}} = {\left( {\sum\limits_{i \in {\mathcal{P}_k}} {{p_i}{{\mathbf{R}}_{il}}}  + {\sigma ^2}{{\mathbf{I}}_N}} \right)^{ - 1}} .
\end{align}
In addition, the distribution of the estimate ${{{\mathbf{\hat h}}}_{kl}}\left[ \lambda  \right]$ and the estimation error ${{{\mathbf{\tilde h}}}_{kl}}\left[ \lambda  \right] = {{\mathbf{h}}_{kl}}\left[ \lambda  \right] - {{{\mathbf{\hat h}}}_{kl}}\left[ \lambda  \right]$ are $\mathcal{C}\mathcal{N}\left( {{\mathbf{0}},{{\mathbf{Q}}_{kl}}} \right)$ and ${{\mathbf{h}}_{kl}}\left[ \lambda  \right] - {{{\mathbf{\hat h}}}_{kl}}\left[ \lambda  \right] \sim \mathcal{C}\mathcal{N}\left( {{\mathbf{0}},{{\mathbf{R}}_{kl}}-{{\mathbf{Q}}_{kl}}} \right)$, where
\begin{align}\label{Qkl}
{{\mathbf{Q}}_{kl}} = {p_k}{e^{ - \left( {\lambda  - {t_k}} \right)\left( {\sigma _{{\text{ap}}}^2 + \sigma _{{\text{ue}}}^2} \right)}}{{\mathbf{R}}_{kl}}{{\mathbf{\Psi }}_{kl}}{{\mathbf{R}}_{kl}} .
\end{align}
Moreover, to simplify the notation, we define
\begin{align}
{{{\mathbf{\bar Q}}}_{kil}} = \sqrt {{p_k}{p_i}} {e^{ - \left( {\lambda  - {t_k}} \right)\left( {\sigma _{{\text{ap}}}^2 + \sigma _{{\text{ue}}}^2} \right)}}{{\mathbf{R}}_{il}}{{\mathbf{\Psi }}_{kl}}{{\mathbf{R}}_{kl}} .
\end{align}

\begin{figure}[t]
\centering
\includegraphics[scale=0.5]{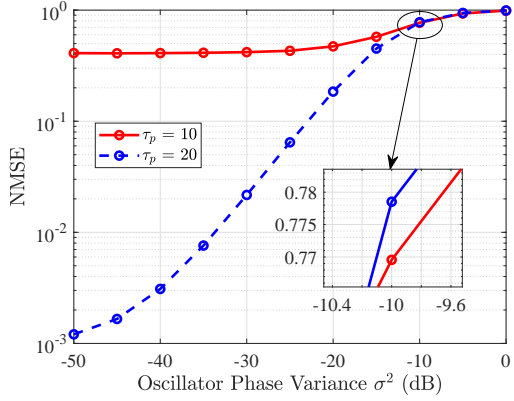}
\caption{NMSE of MMSE channel estimation with different degrees of oscillator phase at the $\text{SNR} = 30$ dB ($L=100$, $K=20$, $N=2$, $\sigma_\text{ap}^2=\sigma_\text{ue}^2=\sigma^2$).} \vspace{-4mm}
\label{figure1}
\end{figure}

\begin{rem}
To compare the estimation quality under different degrees of asynchronous reception, we utilize the normalized mean square error (NMSE) given as
\begin{align}\label{NMSE}
{\mathrm{NMSE}}{_{kl}} = \frac{{{\mathrm{tr}}\left( {{{\mathbf{R}}_{kl}} - {{\mathbf{Q}}_{kl}}} \right)}}{{{\mathrm{tr}}\left( {{{\mathbf{R}}_{kl}}} \right)}}
\end{align}
which is a suitable metric to measure the relative estimation error. Note that ${{{\mathbf{\hat h}}}_{kl}}\left[ \lambda  \right]$ and ${{{\mathbf{\tilde h}}}_{kl}}\left[ \lambda  \right]$ are independent due to the properties of MMSE estimation. Therefore, the values of NMSE are between 0 and 1, which denote perfect and extremely impaired estimation, respectively.
\end{rem}
As illustrated in Fig.~\ref{figure1}, we plot the NMSE of MMSE channel estimation with different degrees of oscillator phase at the $\text{SNR} = 30$ dB. It is clear that the MMSE estimation quality gets worse with the increasing of the oscillator phase variance. The reason is that oscillator phase destroys the orthogonality of the received pilot signal. We also can find that the slope of the NMSE change will be larger when the pilots are fully sufficient. In some extreme asynchronous cases, using non-orthogonal pilots can get more accurate estimation. This result can be explained by \eqref{Qkl} and \eqref{NMSE} as that pilot contamination makes the value of NMSE worse and change slowly.

\section{Downlink Data Transmission}\label{se:performance}

In this part, considering coherent and non-coherent downlink transmission, we utilize both delay phase used (DU) and delay phase forgotten (DF) maximum-ratio (MR) precoding to derive the closed-form SE performance expressions for quantifying the asynchronous effect.

\subsection{Coherent Downlink Transmission}

\newcounter{mytempeqncnt}
\begin{figure*}[t!]
\normalsize
\setcounter{mytempeqncnt}{1}
\setcounter{equation}{23}
\begin{align}\label{SINR_CF_NC}
{\text{SIN}}{{\text{R}}^{\text{nc,du/df}}_k}\left[ n \right] = \frac{{ \eta _n^{{\text{ap}}}\eta _n^{{\text{ue}}} {p_{\text{d}}} \sum\limits_{l = 1}^L {{\mu _l}} {{\left| {{\text{tr}}\left( {{{\mathbf{Q}}_{kl}}} \right)} \right|}^2}}}{{{p_{\text{d}}}\sum\limits_{i = 1}^K {\sum\limits_{l = 1}^L {{\mu _l}} {\text{tr}}\left( {{{\mathbf{Q}}_{il}}{{\mathbf{R}}_{kl}}} \right)} { + }{p_{\text{d}}}\sum\limits_{i \in {\mathcal{P}_k}}^K {\sum\limits_{l = 1}^L {{\mu _l}} {{\left| {{\text{tr}}\left( {{{{\mathbf{\bar Q}}}_{kil}}} \right)} \right|}^2}}  - \eta _n^{{\text{ap}}}\eta _n^{{\text{ue}}} {p_{\text{d}}} \sum\limits_{l = 1}^L {{\mu _l}} {{\left| {{\text{tr}}\left( {{{\mathbf{Q}}_{kl}}} \right)} \right|}^2} + {\sigma ^2_\text{d}}}}
\end{align}
\setcounter{equation}{14}
\hrulefill
\end{figure*}

It is assumed that coherent joint transmission is used for the downlink transmission in the CF massive MIMO, which means that each AP sends the same data symbol to each UE as the other APs. With the linear precoder ${{\mathbf{v}}_{kl}}$, the transmitted signal from AP $l$ at the $n$th time instant is given by
\begin{align}
{{\mathbf{x}}_l}\left[ n \right] = \sqrt {{p_{\text{d}}}{\mu _l}} \sum\limits_{i = 1}^K {{{\mathbf{v}}_{il}}{s_i}\left[ n \right]}
\end{align}
where ${s_i}\left[ n \right]$ represents the universal symbol sent to UE $i$ for all APs, and $p_{\text{d}}$ is the maximum downlink transmission power for one AP. In addition, $\mu _l$ is the normalization parameter for the precoding and it is given by
\begin{align}\label{mu}
{\mu _l}{\text{ = }}{1}/{{\sum\limits_{i = 1}^K {\mathbb{E}\left\{ {{{\left\| {{{\mathbf{v}}_{il}}} \right\|}^2}} \right\}} }}.
\end{align}
The received signal of the $k$th UE at the $n$th time instant is
\begin{align}\label{resignal}
y_k^{{\text{co}}}\left[ n \right] &= \sqrt {{p_{\text{d}}}} \sum\limits_{l = 1}^L {{\mathbf{g}}_{kl}^{\text{H}}\left[ n \right]\sqrt {{\mu _l}} {{\mathbf{v}}_{kl}}{s_k}\left[ n \right]}  \notag\\
&+ \sqrt {{p_{\text{d}}}} \sum\limits_{i \ne k}^K {\sum\limits_{l = 1}^L {{\mathbf{g}}_{kl}^{\text{H}}\left[ n \right]\sqrt {{\mu _l}} {{\mathbf{v}}_{il}}{s_i}\left[ n \right]} }  + {w_k}\left[ n \right]
\end{align}
where ${w_k}\!\left[ n \right] \!\sim\! \mathcal{C}\mathcal{N}\!\left( {0,\sigma _{\text{d}}^2} \right)$ denotes the receiver noise at UE $k$.

\begin{thm}\label{thm1}
With the help of the channel estimate \eqref{hhat} and the received signal in \eqref{resignal}, using the DU-MR precoding ${{\mathbf{v}}_{kl}} = {\theta _{kl}}{{{\mathbf{\hat h}}}_{kl}}\left[ \lambda  \right]$, the downlink capacity of UE k is lower bounded as
\begin{align}
{\mathrm{SE}}_k^{{\mathrm{co,du}}} = \frac{1}{{{\tau _c}}}\sum\limits_{n = \lambda }^{{\tau _c}} {{{\log }_2}\left( {1 + {\mathrm{SINR}}_k^{{\mathrm{co,du}}}\left[ n \right]} \right)}
\end{align}
with ${\mathrm{SINR}}_k^{{\mathrm{co,du}}}\left[ n \right]$ is given by
\begin{align}\label{SINR_CF}
{\mathrm{SINR}}_k^{{\mathrm{co}},{\mathrm{du}}}\left[ n \right] = \frac{{ \eta _n^{{\mathrm{ap}}}\eta _n^{{\mathrm{ue}}} {p_{\mathrm{d}}}{{\left| {\sum\limits_{l = 1}^L {\sqrt {{\mu _l}} {\mathrm{tr}}\left( {{{\mathbf{Q}}_{kl}}} \right)} } \right|}^2}}}{{\Xi _k^{{\mathrm{co}},{\mathrm{du}}}\left[ n \right] + \sigma _{\mathrm{d}}^2}}
\end{align}
where ${\Xi _k^{{\mathrm{co}},{\mathrm{du}}}\left[ n \right]}$ is
\begin{align}
   &{p_{\mathrm{d}}}\!\sum\limits_{i = 1}^K \!{\sum\limits_{l = 1}^L \!{{\mu _l}{\mathrm{tr}}\!\left( {{{\mathbf{Q}}_{il}}{{\mathbf{R}}_{kl}}} \right)}  \!+\! \left( {1 \!-\! \eta _n^{{\mathrm{ap}}} } \right){p_{\mathrm{d}}}\!\sum\limits_{i \in {\mathcal{P}_k}}^K \! {\sum\limits_{l = 1}^L \!{{\mu _l}{{\left| {{\mathrm{tr}}\!\left( {{{{\mathbf{\bar Q}}}_{kil}}} \right)} \right|}^2}} }  }  \notag \\
   &\!+\!\eta _n^{{\mathrm{ap}}}{p_{\mathrm{d}}}\!\!\sum\limits_{i \in {\mathcal{P}_k}}^K \!{{{\left|\! {\sum\limits_{l = 1}^L \!{\sqrt {{\mu _l}} {\mathrm{tr}}\!\left( {{{{\mathbf{\bar Q}}}_{kil}}} \right)} } \!\right|}^2}} \!\!\!-\! \eta _n^{{\mathrm{ap}}}\eta _n^{{\mathrm{ue}}}{p_{\mathrm{d}}}{\left|\! {\sum\limits_{l = 1}^L \!{\sqrt {{\mu _l}} {\mathrm{tr}}\!\left( {{{\mathbf{Q}}_{kl}}} \right)} } \!\right|^2} \!. \notag
\end{align}
Note that we have $\eta _n^{{\mathrm{ap}}} \triangleq {e^{ - \left( {n - \lambda } \right)\sigma _{{\mathrm{ap}}}^2}}$ and $\eta _n^{{\mathrm{ue}}} \triangleq {e^{ - \left( {n - \lambda } \right)\sigma _{{\mathrm{ue}}}^2}}$.
\end{thm}
\begin{IEEEproof}
See Appendix A.
\end{IEEEproof}
\begin{rem}\label{appro}
Only keeping oscillator phase, the approximate SINR expression \eqref{SINR_CF} under which the number of antennas tends to infinity $\left(LN \!\!\to\!\! \infty \right)$ can be derived as $1/\left( {1/\left( {\eta _n^{{\mathrm{ap}}}\eta _n^{{\mathrm{ue}}}} \right) \!+\! 1/\left( {\eta _n^{{\mathrm{ue}}}} \right) \!+\! a} \right)$, where $a$ is a constant. It is clear that the SINR decreases as $\sigma_\mathrm{ap}^2$ and $\sigma_\mathrm{ue}^2$ increase, and $\sigma_\mathrm{ue}^2$ has a larger effect on the SINR compared with $\sigma_\mathrm{ap}^2$.
\end{rem}
From \eqref{SINR_CF}, we can find that the delay phase has no effect on the SINR expression using DU-MR precoding. To study and characterize the influence of delay phase on the system, we investigate the SE performance using DF-MR precoding with ${{\mathbf{v}}_{kl}} = {{{\mathbf{\hat h}}}_{kl}}\left[ \lambda  \right]$. Following similar steps in Theorem~\ref{thm1}, we obtain the ${\text{SINR}}_k^{{\text{co}},{\text{df}}}\left[ n \right]$ as
\begin{align}\label{SINR_CF_DU}
{\text{SINR}}_k^{{\text{co}},{\text{df}}}\left[ n \right] = \frac{{ \eta _n^{{\mathrm{ap}}}\eta _n^{{\mathrm{ue}}} {p_{\text{d}}}{{\left| {\sum\limits_{l = 1}^L {{\theta _{kl}^*}\sqrt {{\mu _l}} {\text{tr}}\left( {{{\mathbf{Q}}_{kl}}} \right)} } \right|}^2}}}{{\Xi _k^{{\text{co}},{\text{df}}}\left[ n \right] + \sigma _{\text{d}}^2}}
\end{align}
where ${\Xi _k^{{\text{co}},{\text{df}}}\!\left[ n \right]}$ is given by
\begin{align}
   &{p_{\text{d}}}\!\sum\limits_{i = 1}^K \!{\sum\limits_{l = 1}^L \!{{\mu _l}{\text{tr}}\!\left( {{{\mathbf{Q}}_{il}}{{\mathbf{R}}_{kl}}} \right)} } \!+\! \left( {1 \!-\! \eta _n^{{\mathrm{ap}}}} \right){p_{\text{d}}}\!\sum\limits_{i \in {\mathcal{P}_k}}^K \!{\sum\limits_{l = 1}^L \!{{\mu _l}{{\left| {{\text{tr}}\!\left( {{{{\mathbf{\bar Q}}}_{kil}}} \right)} \right|}^2}} } \notag \\
   &\!\!\!+\! \eta _n^{{\mathrm{ap}}}\!{p_{\text{d}}}\!\!\sum\limits_{i \in {\mathcal{P}_k}}^K \!{{{\left|\! {\sum\limits_{l = 1}^L \!{{\theta _{il}^*}\!\sqrt {{\mu _l}} {\text{tr}}\!\left(\! {{{{\mathbf{\bar Q}}}_{kil}}} \!\right)} } \!\right|}^2}} \!\!\!\!-\! \eta _n^{{\mathrm{ap}}}\!\eta _n^{{\mathrm{ue}}}\!{p_{\text{d}}}{\left|\! {\sum\limits_{l = 1}^L \!{{\theta _{kl}^*}\!\sqrt {{\mu _l}} {\text{tr}}\!\left(\! {{{\mathbf{Q}}_{kl}}} \!\right)} } \!\right|^2} \!. \notag
\end{align}

\subsection{Non-coherent Downlink Transmission}

For alleviating the phase-synchronization requirements of APs imposed by coherent transmission, we apply non-coherent joint transmission in downlink CF massive MIMO, which implies that each AP is able to send data to each UE but does so using a different data symbol than the other APs.
Then, adopting the linear precoder ${{\mathbf{v}}_{kl}}$, the transmitted signal from AP $l$ at time instant $n$ is expressed as
\begin{align}
{{\mathbf{x}}_l}\left[ n \right] = \sqrt {{p_{\text{d}}}{\mu _l}} \sum\limits_{i = 1}^K {{{\mathbf{v}}_{il}}{s_{il}}\left[ n \right]}
\end{align}
where ${{s_{il}}\left[ n \right]} \sim \mathcal{C}\mathcal{N}\left( {0,1} \right)$ denotes the symbol sent to UE $i$ which is different for all APs, and $p_\text{d}$ represents the maximum downlink transmission power for one AP. It is equal to \eqref{mu}, ${{\mu _l}}$ is the precoding normalization parameter selected to meet the downlink power constraint as $\mathbb{E}\left\{ {\left\| {{{\mathbf{x}}_l}\left[ n \right]} \right\|} \right\} \leqslant {p_{\text{d}}}$. Then, the received signal of the $k$th UE at the $n$th time instant is
\begin{align}\label{resignal1}
y_k^{{\text{nc}}}\left[ n \right] &= \sqrt {{p_{\text{d}}}} \sum\limits_{l = 1}^L {{\mathbf{g}}_{kl}^{\text{H}}\left[ n \right]\sqrt {{\mu _l}} {{\mathbf{v}}_{kl}}{s_{kl}}\left[ n \right]}  \notag\\
&+ \sqrt {{p_{\text{d}}}} \sum\limits_{i \ne k}^K {\sum\limits_{l = 1}^L {{\mathbf{g}}_{kl}^{\text{H}}\left[ n \right]\sqrt {{\mu _l}} {{\mathbf{v}}_{il}}{s_{il}}\left[ n \right]} }  + {w_k}\left[ n \right] .
\end{align}
Note that the $k$th UE needs to employ successive interference cancellation after receiving the signals from all $L$ APs in order to detect the signals sent by the different APs \cite{9322468}. Specifically, the UE first detects the signal received from the first AP, and the remaining signal is treated as interference. By that analogy, the UE detects the signal transmitted by the $l$th AP, and considers the signal sent from the $(l + 1)$th AP to the $L$th AP as interference, thereby detecting the signal ${s_{kl}}\left[ n \right]$.

\begin{thm}
Based on the received signal \eqref{resignal1} and successive interference cancellation method, and using the DU-MR precoding ${{\mathbf{v}}_{kl}} = {\theta _{kl}}{{{\mathbf{\hat h}}}_{kl}}\left[ \lambda  \right]$, the downlink capacity of UE $k$ is lower bounded as
\begin{align}
{\mathrm{SE}}_k^{{\mathrm{nc,du}}} = \frac{1}{{{\tau _c}}}\sum\limits_{n = \lambda }^{{\tau _c}} {{{\log }_2}\left( {1 + {\mathrm{SINR}}_k^{{\mathrm{nc,du}}}\left[ n \right]} \right)}
\end{align}
where ${\mathrm{SINR}}_k^{{\mathrm{nc,du}}}\left[ n \right]$ is given by \eqref{SINR_CF_NC} at the top of this page.
\end{thm}
\begin{IEEEproof}
Follow the similar steps in Theorem \ref{thm1}.
\end{IEEEproof}

\begin{rem}
Note that if the DF-MR precoding is used, we can also derive the same SINR expression as \eqref{SINR_CF_NC} at the top of this page. Therefore, we conclude that the non-coherent transmission can effectively overcome the influence of asynchronous reception, and even eliminate the influence of delay phase, at the expense of poor SE.
\end{rem}
\setcounter{equation}{24}

\section{Numerical Results and Discussion}\label{se:numerical}

We use the three-slope propagation model in a simulation setup in which $K$ UEs and $L$ APs are uniformly and independently distributed within a square of size $500$ m $\times$ $500$ m \cite{Ngo2017Cell}.
It is assumed that the bandwidth is $B\!=\!20$ MHz and the carrier frequency is $f_c\!=\!2$ GHz. In addition, both the pilot and data transmission power are $p \!= p_\text{d} = \!23$ dBm. Moreover, the noise power is $\sigma^2\!=\!-96$ dBm, and the coherence block has $\tau_c \!=\! 200$ channel uses \cite{bjornson2019making}.

\begin{figure}[t]
\centering
\includegraphics[scale=0.5]{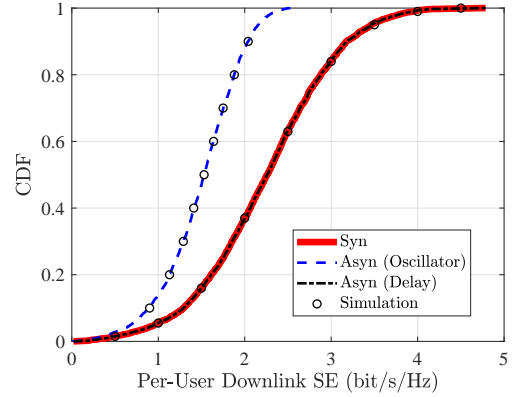}
\caption{CDF of per-user downlink SE for CF massive MIMO systems with DU-MR combining under asynchronous reception ($L=100$, $K=20$, $N=2$, $\tau_p = 10$).} \vspace{-4mm}
\label{figure2}
\end{figure}

Taking into account oscillator and delay phase respectively, Fig.~\ref{figure2} shows the CDF of per-user downlink SE for CF massive MIMO with coherent transmission and DU-MR precoding. It is found that the oscillator phase has a bad effect on the SE performance, because the coherent data transmission gain of CF massive MIMO systems is destroyed by the asynchronous reception. Besides, due to the delay phase is perfectly known and used for DU-MR precoding, the coherent transmission and SE performance is not effected. Note that this result can also be found from \eqref{SINR_CF}. Moreover, simulation results verify the correctness of our derived closed-form SE expressions.

\begin{figure}[t]
\centering
\includegraphics[scale=0.5]{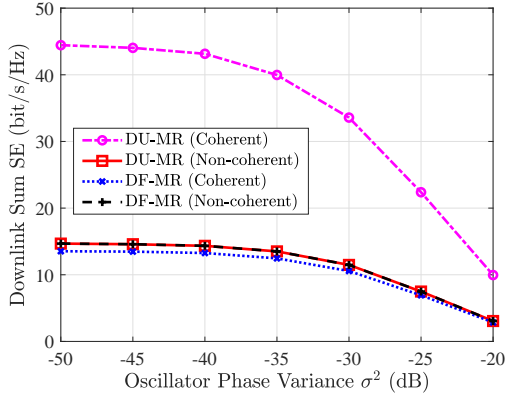}
\caption{Downlink sum SE for CF massive MIMO systems with different degrees of oscillator phase ($L=100$, $K=20$, $N=2$, $\tau_p = 10$, $\sigma_\text{ap}^2=\sigma_\text{ue}^2=\sigma^2$).} \vspace{-4mm}
\label{figure3}
\end{figure}

Figure~\ref{figure3} compares the downlink sum SE of CF massive MIMO systems against degrees of oscillator phase under different data transmission and MR precoding modes. It is clear that the case on coherent transmission and  DU-MR drops fastest with the oscillator phase variance increases. The reason is that coherent transmission requires tight synchronization. For example, in the case with DF-MR, the existence of delay phase makes coherent transmission impossible. Note that the propagation distance offset that causes delay phase is often several hundred times the wavelength ($\lambda$), which affects the system much more than the oscillator phase. Therefore, in the precoding design of the distributed architecture, it is necessary to obtain the accurate delay phase by some method such as positioning technology. In addition, we also find that the case of non-coherent transmission is not effected by the delay phase, and is better than coherent transmission when the delay phase is unknown.

\begin{figure}[t]
\centering
\includegraphics[scale=0.5]{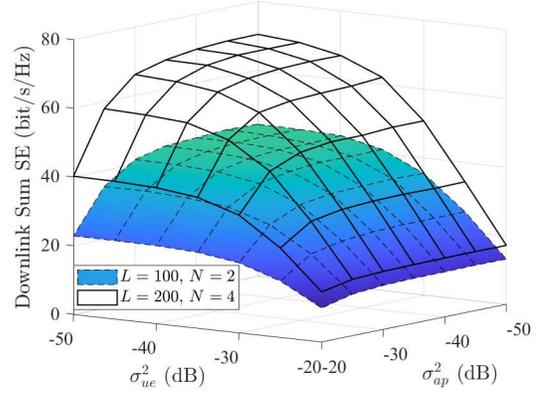}
\caption{Downlink sum SE for CF massive MIMO systems against different degrees of oscillator phase at AP and UE ($K=20$, $\tau_p = 10$).} \vspace{-4mm}
\label{figure4}
\end{figure}

The downlink sum SE for CF massive MIMO systems with DU-MR precoding is shown in Fig.~\ref{figure4}, as a decreasing function of degrees of oscillator phase both at AP and UE. We notice that the oscillator phase of UE has a larger bad effect on SE performance than that of AP. For instance, at the case $L=100$, $N=2$, only increasing $\sigma^2_\text{ap}$ from $-50$ dB to $-20$ dB will results in 48\% SE loss, but the same operation on $\sigma^2_\text{ue}$ will cause 70\% SE loss. It is worthy noting that this different degree of influence becomes more pronounced as the number of antennas increases, which also can be obtained by Remark~\ref{appro}. Moreover, we also find that varying the number of antennas from $L=100$, $N=2$ to $L=200$, $N=4$ has 30\% SE gain at the case $\sigma^2_\text{ap}=-50$ dB, $\sigma^2_\text{ue}=-20$ dB, and leads to 76\% SE gain at the case $\sigma^2_\text{ap}=-20$ dB, $\sigma^2_\text{ue}=-50$ dB. The reason is that the more antennas promise higher antenna array gains offering more degrees of freedom to compensate the negative impact caused by the oscillator phase at the AP.

\newcounter{mytempeqncnt1}
\begin{figure*}[t!]
\normalsize
\setcounter{mytempeqncnt}{1}
\setcounter{equation}{36}
\begin{align}\label{INTc}
{\text{IN}}{{\text{T}}_i}\left[ n \right] = \sum\limits_{l = 1}^L {{\mu _l}{\text{tr}}\left( {{{\mathbf{Q}}_{il}}{{\mathbf{R}}_{kl}}} \right){\text{ + }}\left\{ {\begin{array}{*{20}{c}}
  {\left( {1 - {e^{ - \left( {n - \lambda } \right)\left( {\sigma _{{\text{ap}}}^2} \right)}}} \right)\sum\limits_{l = 1}^L {{\mu _l}{{\left| {{\text{tr}}\left( {{{{\mathbf{\bar Q}}}_{kil}}} \right)} \right|}^2}}  + {e^{ - \left( {n - \lambda } \right)\left( {\sigma _{{\text{ap}}}^2} \right)}}{{\left| {\sum\limits_{l = 1}^L {\sqrt {{\mu _l}} {\text{tr}}\left( {{{{\mathbf{\bar Q}}}_{kil}}} \right)} } \right|}^2},i \in {\mathcal{P}_k}} \\
  {0,i \notin {\mathcal{P}_k}}
\end{array}} \right.}
\end{align}
\setcounter{equation}{24}
\hrulefill
\end{figure*}

\section{Conclusion}\label{se:conclusion}

We investigated the performance of CF massive MIMO systems with asynchronous reception, including both delay and oscillator phases. Taking into account the coherent and non-coherent transmission, we derived novel closed-form SE expressions for CF massive MIMO systems with channel estimation errors caused by phase-asynchronization and pilot contamination. It was shown that asynchronous reception destroys pilot orthogonality and coherent data transmission, resulting in poor system performance. In particular, obtaining an accurate delay phase is important for CF massive MIMO systems to realize coherent transmission. Moreover, it is interesting that the oscillator phase of UEs has a larger effect on SE performance than that of APs, and increasing the number of antennas can significantly reduce the influence of the oscillator phase at APs.
In future work, possible solutions to solve asynchronous problems in CF massive MIMO systems will be investigated, e.g., the small UE-centric cluster formation \cite{zhang2021local}, and the rate-splitting/NOMA against interference \cite{7434643}.

\begin{appendices}
\section{Proof of Theorem 1}

Each UE is supposed to be aware of the channel statistics and the signal detection is performed with channel distribution information. Based on this assumption, we can derive the use-and-then-forget capacity bound with SINR is given by
\begin{align}
{\text{SIN}}{{\text{R}}_k}\left[ n \right] = \frac{{p{{\left| {\sum\limits_{l = 1}^L {{\text{D}}{{\text{S}}_{kl}}\left[ n \right]} } \right|}^2}}}{{p\sum\limits_{i = 1}^K {{\text{IN}}{{\text{T}}_i}\left[ n \right]}  - p{{\left| {\sum\limits_{l = 1}^L {{\text{D}}{{\text{S}}_{kl}}\left[ n \right]} } \right|}^2} + {\sigma ^2}}}
\end{align}
where ${{\text{D}}{{\text{S}}_{kl}}\left[ n \right]}$ denotes the desired signal and ${{\text{IN}}{{\text{T}}_i}\left[ n \right]}$ is the interference from other UEs\footnote{Note that $\text{INT}_k$ denotes the instantaneous component of UE $k$, which is subtracted by its statistical component $\text{DS}_k$ to obtain the beamforming gain uncertainty as the residual interference part.}. While the DU-MR precoding ${{\mathbf{v}}_{kl}} = {\theta _{kl}}{{{\mathbf{\hat h}}}_{kl}}\left[ \lambda  \right]$ is used, they can be expressed as
\begin{align}
\label{DS} {{\text{D}}{{\text{S}}_{kl}}\left[ n \right]} &= {\mathbb{E}\left\{ {{\mathbf{g}}_{kl}^{\text{H}}\left[ n \right]\sqrt {{\mu _l}} {{\theta _{kl}}{{{\mathbf{\hat h}}}_{kl}}\left[ \lambda  \right]}} \right\}} \\
{{\text{IN}}{{\text{T}}_i}\left[ n \right]} &= {\mathbb{E}\left\{ {{{\left| {\sum\limits_{l = 1}^L {{\mathbf{g}}_{kl}^{\text{H}}\left[ n \right]\sqrt {{\mu _l}} {{\theta _{il}}{{{\mathbf{\hat h}}}_{il}}\left[ \lambda  \right]}} } \right|}^2}} \right\}} .
\end{align}
Moreover, the normalization parameter regarding the precoding in \eqref{mu} can be written as $ {\mu _l} = 1/\sum\nolimits_{i = 1}^K {{\text{tr}}\left( {{{\mathbf{Q}}_{il}}} \right)} $.
Submitting \eqref{gkl} into \eqref{DS}, and using the definition and property of $\Theta _{kl}$ in \eqref{theta1} and \eqref{theta2}, we can derive
\begin{align}
   {{\text{D}}{{\text{S}}_{kl}}\left[ n \right]} &= \mathbb{E}\left\{ {\Theta _{kl}^*\left[ n - \lambda \right]} \right\}\sqrt {{\mu _l}} \; \mathbb{E}\left\{ {{\mathbf{h}}_{kl}^{\text{H}}\left[ \lambda  \right]{{{\mathbf{\hat h}}}_{kl}}\left[ \lambda  \right]} \right\} \notag \\
   &= {e^{ - \frac{{n - \lambda }}{2}\left( {\sigma _{{\text{ap}}}^2 + \sigma _{{\text{ue}}}^2} \right)}}\sqrt {{\mu _l}} {\text{tr}}\left( {{{\mathbf{Q}}_{kl}}} \right) .
\end{align}
Moreover, with the help of \cite[Eq. (69)]{9322468}, we have
\begin{align}\label{INT}
  &{\text{IN}}{{\text{T}}_i}\left[ n \right] = \sum\limits_{l = 1}^L {\underbrace {\mathbb{E}\left\{ {{{\left| {{\mathbf{g}}_{kl}^{\text{H}}\left[ n \right]\sqrt {{\mu _l}} {\theta _{il}}{{{\mathbf{\hat h}}}_{il}}\left[ \lambda  \right]} \right|}^2}} \right\}}_{{\Upsilon _1}}}  + \sum\limits_{l = 1}^L {\mathop \sum \limits_{m \ne l}^L }  \notag \\
   &\!\times\! \underbrace {\mathbb{E}\!\left\{\! {{{\left( {{\mathbf{g}}_{kl}^{\text{H}}\!\left[ n \right]\!\sqrt {{\mu _l}} {\theta _{il}}{{{\mathbf{\hat h}}}_{il}}\!\left[ \lambda  \right]} \!\right)}^*}\!\!\left(\! {{\mathbf{g}}_{km}^{\text{H}}\!\left[ n \right]\!\sqrt {{\mu _m}} {\theta _{im}}{{{\mathbf{\hat h}}}_{im}}\!\left[ \lambda  \right]} \right)} \!\right\}}_{{\Upsilon _2}} .
\end{align}
Besides, with the help of \eqref{gkl} and \eqref{theta1}, we derive
\begin{align}\label{gamma1}
  {\Upsilon _1} &= \mathbb{E}\left\{ {{{\left| {{\mathbf{h}}_{kl}^{\text{H}}\left[ \lambda  \right]\Theta _{kl}^*\left[ n - \lambda \right]\theta _{kl}^*\sqrt {{\mu _l}} {\theta _{il}}{{{\mathbf{\hat h}}}_{il}}\left[ \lambda  \right]} \right|}^2}} \right\} \notag \\
   &= {\mu _l}\mathbb{E}\left\{ {{{\left| {{\mathbf{h}}_{kl}^{\text{H}}\left[ \lambda  \right]{{{\mathbf{\hat h}}}_{il}}\left[ \lambda  \right]} \right|}^2}} \right\} .
\end{align}
Following the similar steps of \cite[Eq. (62)]{9322468}, we write \eqref{gamma1} as
\begin{align}\label{gamma11}
{\Upsilon _1} = {\mu _l}{\text{tr}}\left( {{{\mathbf{Q}}_{il}}{{\mathbf{R}}_{kl}}} \right) + \left\{ {\begin{array}{*{20}{c}}
  {{\mu _l}{{\left| {{\text{tr}}\left( {{{{\mathbf{\bar Q}}}_{kil}}} \right)} \right|}^2},i \in {\mathcal{P}_k}} \\
  {0,i \notin \mathcal{P}} .
\end{array}} \right.
\end{align}
By using \eqref{gkl} and \eqref{theta1}, we obtain
\begin{align}\label{gamma2}
  {\Upsilon _2} &= {\theta _{kl}}\theta _{il}^*\theta _{km}^*{\theta _{im}}\sqrt {{\mu _l}{\mu _m}} \;\underbrace {\mathbb{E}\left\{ {{\Theta _{kl}}\left[ n - \lambda \right]\Theta _{km}^*\left[ n - \lambda \right]} \right\}}_{{\Upsilon _3}} \notag \\
   &\times {\underbrace {\mathbb{E}\left\{ {{\mathbf{h}}_{kl}^{\text{H}}\left[ \lambda  \right]{{{\mathbf{\hat h}}}_{il}}\left[ \lambda  \right]} \right\}}_{{\Upsilon _4}}}^*\mathbb{E}\left\{ {{\mathbf{h}}_{km}^{\text{H}}\left[ \lambda  \right]{{{\mathbf{\hat h}}}_{im}}\left[ \lambda  \right]} \right\} .
\end{align}
By utilizing the definition and property of $\Theta_{kl}$ in \eqref{theta1} and \eqref{theta2}, we derive
\begin{align}\label{gamma3}
  &{{\Upsilon _3}} = \mathbb{E}\left\{ {{e^{j\sum\limits_{s = \lambda  + 1}^n {\left( {\delta _k^{{\text{ue}}}\left[ s \right] + \delta _l^{{\text{ap}}}\left[ s \right]} \right)} }}{e^{ - j\sum\limits_{s = \lambda  + 1}^n {\left( {\delta _k^{{\text{ue}}}\left[ s \right] + \delta _m^{{\text{ap}}}\left[ s \right]} \right)} }}} \right\} \notag \\
   & = \mathbb{E}\!\left\{\! {{e^{j\sum\limits_{s = \lambda  + 1}^n {\delta _l^{{\text{ap}}}\left[ s \right]} }}} \!\right\}\mathbb{E}\!\left\{\! {{e^{ - j\sum\limits_{s = \lambda  + 1}^n {\delta _m^{{\text{ap}}}\left[ s \right]} }}} \!\right\} \!=\! {e^{ - \left( {n - \lambda } \right)\sigma _{{\text{ap}}}^2}} .
\end{align}
Besides, based on the properties of MMSE estimation, where ${{{\mathbf{\hat h}}}_{kl}}\left[ \lambda  \right]$ and ${{{\mathbf{\tilde h}}}_{kl}}\left[ \lambda  \right]$ are independent. We have
\begin{align}\label{gamma4}
{\Upsilon _4} = \mathbb{E}\left\{ {{\mathbf{\hat h}}_{kl}^{\text{H}}\left[ \lambda  \right]{{{\mathbf{\hat h}}}_{il}}\left[ \lambda  \right]} \right\} = {\text{tr}}\left( {\mathbb{E}\left\{ {{{{\mathbf{\hat h}}}_{il}}\left[ \lambda  \right]{\mathbf{\hat h}}_{kl}^{\text{H}}\left[ \lambda  \right]} \right\}} \right) .
\end{align}
Submitting \eqref{hhat} into \eqref{gamma4}, we then obtain
\begin{align}\label{gamma44}
{\Upsilon _4} = \left\{ {\begin{array}{*{20}{c}}
  {{\theta _{kl}}\theta _{il}^{\text{*}}{\text{tr}}\left( {{{{\mathbf{\bar Q}}}_{kil}}} \right),i \in {\mathcal{P}_k}} \\
  {0,i \notin {\mathcal{P}_k}} .
\end{array}} \right.
\end{align}
Finally, with the help of \eqref{gamma3} and \eqref{gamma44} and plugging \eqref{gamma11} and \eqref{gamma2} into \eqref{INT}, we derive the ${\text{IN}}{{\text{T}}_i}\left[ n \right]$ as \eqref{INTc} at the top of this page, and this completes the proof.

\end{appendices}

\vspace{0cm}
\bibliographystyle{IEEEtran}
\bibliography{IEEEabrv,Ref}

\end{document}